# Ultra-low lattice thermal conductivity of MgPb$_2$Te- A first principles study

Rajmohan Muthaiah, Jivtesh Garg

School of Aerospace and Mechanical Engineering, University of Oklahoma, Norman,

OK-73019, USA

**Abstract:** Thermoelectric technology is an alternate way to efficiently utilize the energy by converting waste heat into electricity. Thermoelectric requires material with low thermal conductivity to improves its thermoelectric performance. In this work, by solving Boltzmann transport equation based on first principles calculations, we report an ultra-low room temperature thermal conductivity of 2.08 Wm$^{-1}$K$^{-1}$ and 2.9 Wm$^{-1}$K$^{-1}$ along c-axis and a-axis respectively for pure MgPb$_2$Te. To explain this ultra-low thermal conductivity, we analyzed the elastic constants, phonon group velocity, phonon-phonon scattering and contribution from transverse acoustic, longitudinal acoustic and optical phonon branches. We also report the thermal conductivity of MgPb$_2$Te nanostructures. At 50 nm, the room temperature thermal conductivity of MgPb$_2$Te is 0.957 Wm$^{-1}$K$^{-1}$ and 1.459 Wm$^{-1}$K$^{-1}$ along c-axis and a-axis respectively. Ultra-low thermal conductivity unraveled in this work shows MgPb$_2$Te would a promising material for thermoelectric applications.

**Introduction:** Thermoelectric is a clean and renewable energy technology plays an important role in minimizing the green gas emissions by converting waste heat into electricity[1]. On the other hand, thermoelectric coolers can be used to make electronic cooling systems and refrigerators[2-4]. Despite having potential advantages, thermoelectric devices are limited due to its very low efficiency. Hence, improving the thermoelectric efficiency is a major research focus among the researchers. Figure of merit ($ZT = S^2\sigma T/k$)[5] is an estimate of thermoelectric efficiency, where $S$, $\sigma$, $T$, and $k$ denotes the Seebeck coefficient, electrical conductivity, working temperature and thermal conductivity respectfully. Engineering materials with an ultra-low thermal conductivity is critical for thermoelectric applications[6, 7]. Thermal conductivity($k$) of a material can be modulated by controlling the phonon scattering through strain[8-11], nanostructures[12, 13] and point-defects[14] etc., Tellurium based compounds such as PbTe[15, 16], Bi$_2$Te$_3$[17, 18] were primarily reported for thermoelectric applications with its ultra-low lattice thermal conductivity. Recently, magnesium-based compounds such as Mg$_3$Sb$_2$[19, 20], Mg$_3$Bi$_2$[21], CaMg$_2$Sb$_2$[22] and YbMg$_2$Sb$_2$[23] etc., were reported with ultra-low thermal conductivity for thermoelectric applications. In this work, we report an ultra-low thermal conductivity of bulk and nanostructured $MgPb_2Te$. We compared our result with MgTe, PbTe and $Mg_3Sb_2$. At room temperature, thermal conductivity of pure bulk $MgPb_2Te$ (2.08 $Wm^{-1}K^{-1}$ along the c-axis and 2.9 $Wm^{-1}K^{-1}$ along the a-axis) is comparable to thermal conductivity of PbTe(2.1 $Wm^{-1}K^{-1}$)[24] and $Mg_3Sb_2$(1.5 $Wm^{-1}K^{-1}$)[25]. To explain this ultra-low thermal conductivity, we analyzed the phonon group velocity, phonon-phonon scattering and contribution from transverse acoustic, longitudinal acoustic and optical phonon branches. We also report thermal conductivity of the nanostructured $MgPb_2Te$ computed by imposing boundary scattering[26]. Room temperature thermal conductivity for the nanostructured (L=100 nm) $MgPb_2Te$ is 1.21 $Wm^{-1}K^{-1}$ and 1.83 $Wm^{-1}K^{-1}$ along the c-axis and a-axis respectively. These ultra-low thermal conductivity of $MgPb_2Te$ shows the promising nature of applications in thermoelectric applications. This work provides an avenue to further explore this $MgPb_2Te$ with point-defects and alloying etc.,

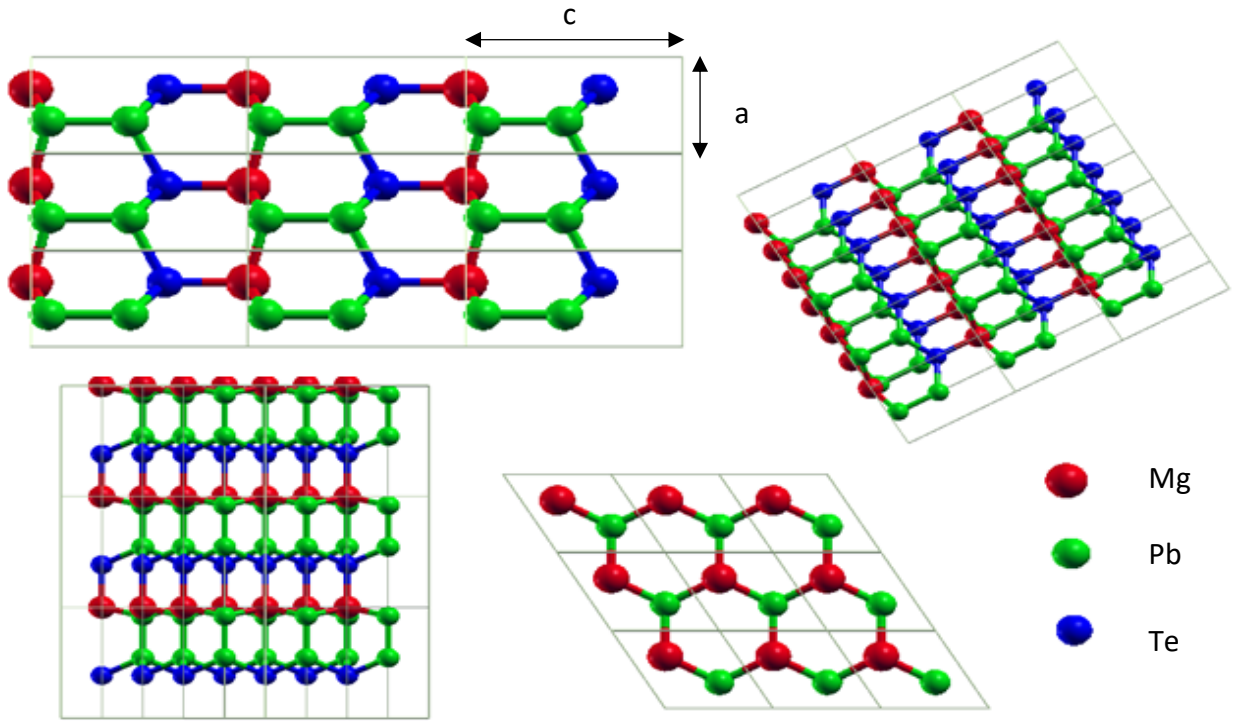

Figure 1: Atomic arrangements of $MgPb_2Te$ with 3 x 3 x 3 supercell with lattice constant a= 8.722 bohr and c/a= 1.6475. Red, green and blue represents the Magnesium(Mg), Lead(Pb) and Tellurium(TE)

**Computational Methods:** All the first principles calculations were carried out using QUANTUM-ESPRESSO[27] package. Norm-conserving pseudopotentials in the local-density approximation (LDA) were used with a plane-wave cut-off of 90 Ry. The geometry of the hexagonal MgPb$_2$Te with 4 atoms unit cell is optimized until the forces on all the atoms are less than $10^{-5}$ ev Å$^{-1}$. Monkhorst[28] $k$-point mesh of 12 x 12 x 8 with gaussian smearing of 0.02 Ry was used to integrate the electronic properties. Elastic constants were calculated using QUANTUM ESPRESSO thermo_pw package. Voigt-Ruess-Hill approximation[29] is used to calculate the bulk modulus(B), Youngs modulus(E) and Shear modulus(G). Thermal conductivity of MgPb$_2$Te was calculated by solving the following phonon Boltzmann transport equation (PBTE)[30],

$$k_\alpha = \frac{\hbar^2}{N\Omega k_b T^2} \sum_\lambda v_{\alpha\lambda}^2 \omega_\lambda^2 \bar{n}_\lambda (\bar{n}_\lambda + 1)\tau_\lambda \tag{1}$$

where $\alpha$, N, $\hbar$, $\Omega$, $k_b$, T, are the cartesian direction, size of the q mesh used for summation, Planck constant, unit cell volume, Boltzmann constant, and absolute temperature respectively. $\omega_\lambda, \bar{n}_\lambda, \tau_\lambda$ and $v_{\alpha\lambda}$ (= $\partial\omega_\lambda/\partial q$) are the phonon frequency, equilibrium Bose-Einstein population, phonon lifetime (inverse of the scattering rate) and group velocity along cartesian direction $\alpha$, respectively of a phonon mode $\lambda$. $\omega_\lambda, \bar{n}_\lambda$ and $v_{\alpha\lambda}$ can be calculated from the knowledge of 2$^{nd}$ order force constants. Phonon lifetime can be calculated based on the 2$^{nd}$ and 3$^{rd}$ order force constants. Harmonic (2$^{nd}$ order) and anharmonic (3$^{rd}$ order) force constants (IFCs), needed for $k$ prediction, are derived from density-functional theory (DFT)[31, 32]. 8 x 8 x 6 $q$-grid was used to compute dynamical matrices and the 2nd order IFCs. 3rd order IFCs were computed on a 4 x 4 x 3 q-point grid using D3Q[33-35] package. Acoustic sum rules were imposed on both 2nd and 3rd order force constants. Phonon linewidth and $k$ calculations were carried out in QUANTUM ESPRESSO thermal2 code with 30 x 30 x 20 $q$-mesh and 0.05 cm$^{-1}$ smearing until the Δk values are less than $1.0e^{-5}$. Iterative solutions were converged after 5 iterations.

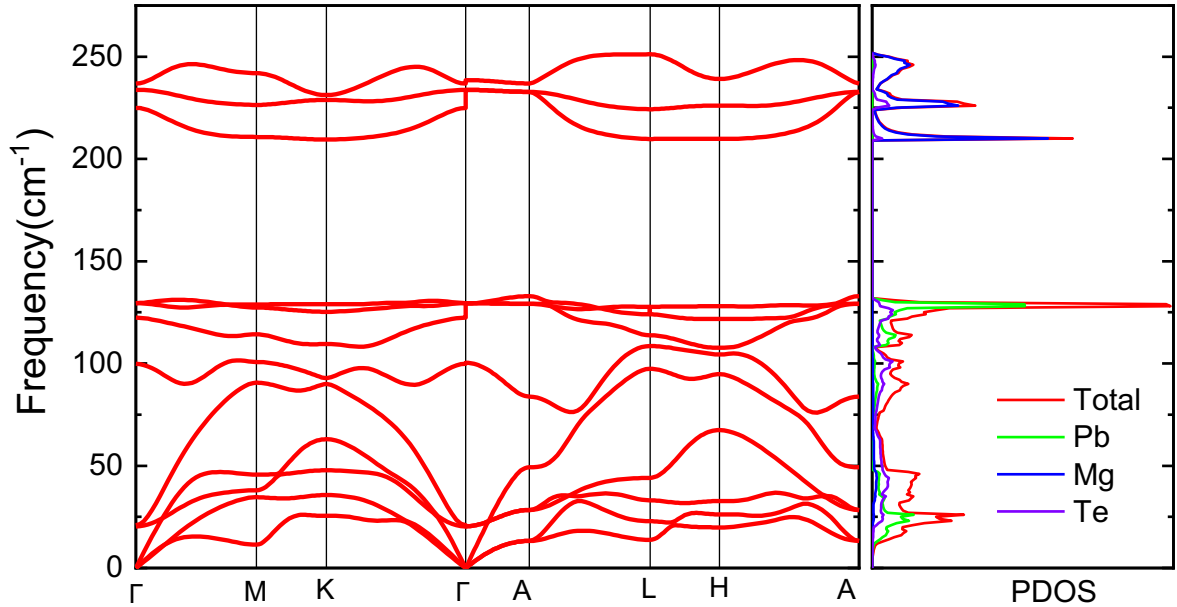

Figure 2: Phonon dispersion and phonon density of states MgPb2Te with lattice constants a= 8.722 bohr and c/a= 1.6475.

**Results and Discussion:** MgPb$_2$Te with equilibrium lattice constants (a=8.722 bohr and c/a=1.6475) and atomic coordinates are shown in Fig. 1. Atomic positions in the unit cell are Pb (4.3614, 2.518,7.1547), Mg (0.0000, 5.036,14.3143), Te (4.3614,2.518,12.701) and Pb (0.00,5.036,5.3299) bohr units. Elastic constants for the MgPb$_2$Te at 0 GPa are reported in Table 1 and the value are $C_{11}$= 61.19 GPa, $C_{33}$=62.69 GPa, $C_{44}$=10.88 GPa, $C_{66}$=16.86 GPa, $C_{12}$=27.46 GPa, $C_{13}$=19.52 GPa which satisfies the Born stability criteria[36] of $C_{66} = (C_{11}-C_{12})/2$, $C_{11} > C_{12}$, $C_{33}(C_{11}+C_{12}) > 2(C_{13})2$, $C_{44} > 0$, $C_{66} > 0$ and hence the system is mechanically stable. Bulk modulus(B), Youngs modulus(E), Shear modulus(G) and poisson ratio based on Voigt-Ruess-Hill approximations[29] are 35.63 GPa, 39.56 GPa, 15.04 GPa and 0.31462 respectively. Phonon dispersion and phonon density of states for hexagonal MgPb$_2$Te is shown in Fig 2. Since all the phonon modes has a positive frequency, our system is dynamically stable[37]. Frequencies less than

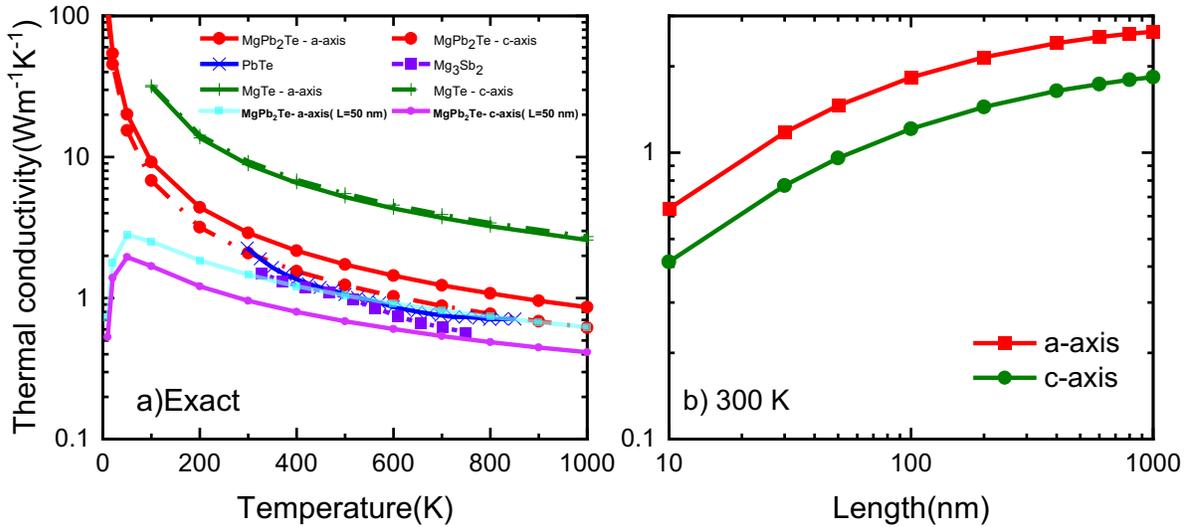

Figure 3a) Temperature dependent thermal conductivity of MgPb$_2$Te along different directions with PbTe, Mg3Sb2 and MgTe b) Length dependence thermal conductivity between 10 nm and 1000 nm at room temperature

130 cm$^{-1}$ is due to the heavy constituent elements such as lead and telluride whereas frequency above 200 cm$^{-1}$ is due to the light element magnesium. Since MgPb$_2$Te satisfied both mechanical and dynamical stability, we proceeded to thermal conductivity calculations.

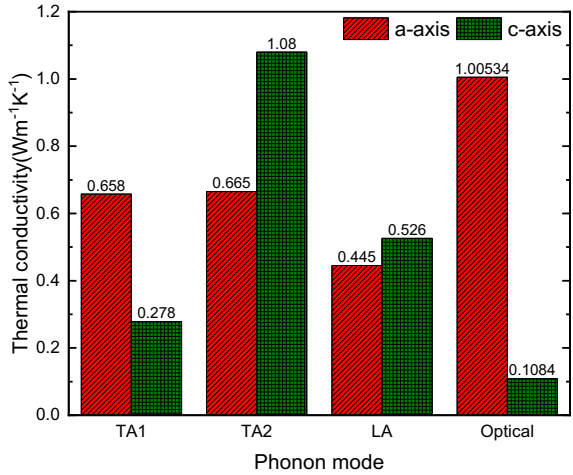

Figure 4 : Mode contribution thermal conductivity of TA$_1$, TA$_2$, LA and optical phonon modes for MgPb$_2$Te

Temperature dependent lattice thermal conductivity($k$) of isotopically pure MgPb$_2$Te along a-axis and c-axis is shown in Fig 3a. First principles computations reveal that, iterative(exact) room temperature thermal conductivity of bulk MgPb$_2$Te along a-axis and c-axis are 2.9 Wm$^{-1}$K$^{-1}$ and 2.08 Wm$^{-1}$K$^{-1}$ respectively. Thermal conductivity computed within the Single mode relaxation time approximation (SMA) are just 4.8% lower than that of the full iterative solution. Our reported values are comparable to the state of the art thermoelectric materials such as PbTe(2.1 Wm$^{-1}$K$^{-1}$)[24] and Mg$_3$Sb$_2$(1.5 Wm$^{-1}$K$^{-1}$)[25]. At 300 K, $k$ of MgPb$_2$Te with sample size between 10 nm and 1000 nm is shown in Fig 3b. Ultra-low thermal conductivity of 0.957 Wm$^{-1}$K$^{-1}$(c-axis) and 1.459 Wm$^{-1}$K$^{-1}$(a-axis) for the nanometer length scales of L = 50 nm suggesting potential applications of MgPb$_2$Te in thermoelectric applications.

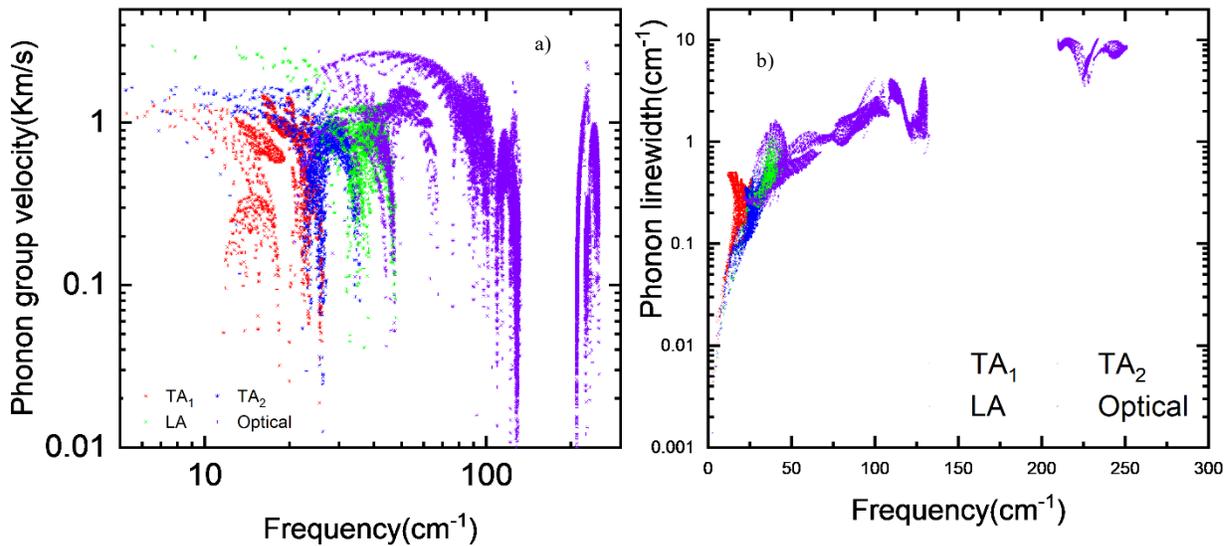

Figure 5a) Phonon group velocity and b) Phonon linewidth (inverse of lifetime) for MgPb$_2$Te

This ultra-low thermal conductivity is mainly attributed with its very low phonon group velocity of both transverse acoustic and longitudinal acoustic phonons which are usually the major heat carrying phonons and optical phonons being the scattering channels for the acoustic phonons[38]. These very low phonon group velocity ($= \partial\omega_\lambda/\partial q$) is due to the presence of heavy elements such as lead and telluride and weak bonding (as we can observe it from the elastic constants). To elucidate this ultra-low thermal conductivity, we also analyzed the mode contribution thermal conductivity of transverse acoustic (TA), longitudinal acoustic (LA) and optical phonon modes as well as phonon group velocities and phonon linewidths (inverse of phonon lifetime). Mode contribution from each phonon branch at SMA is shown in Fig 4. It is interesting to note that, along the a-axis, optical phonons contribute to ~37% of its overall thermal conductivity due to its high phonon group velocity of low frequency optical phonons which we can observe it from phonon dispersion (Γ-X in Fig 2b) and phonon group velocity (Fig 5a). Whereas, along the c-axis, optical phonon frequency contribution is ~5% to its overall thermal conductivity because there is no such high phonon group velocity optical phonons along the c-axis (Γ-A). This causes an anisotropy in the system. TA phonons contributes to 47.5% and 68.6% of its overall thermal conductivity along a-axis and c-axis respectively. Likewise, LA phonons contributes to 15.5% and 26.4% of its overall thermal conductivity along a-axis and c-axis respectively.

Figure 5a and b represents the phonon group velocities and phonon linewidth (inverse of phonon lifetime) of different phonon branches (TA, LA and optical phonons) for MgPb$_2$Te. Again here, we can observe that some of the low frequency optical phonons (< 75 cm$^{-1}$) has considerable phonon group velocity and phonon lifetime as that of the acoustic phonons and causing the considerable contribution to its overall thermal conductivity.

**CONCLUSION:** In summary, by solving phonon Boltzmann transport equation coupled with first principles calculations, we analyzed the lattice thermal conductivity of MgPb$_2$Te. Our first principles computations report an ultra-low thermal conductivity of 2.08 Wm$^{-1}$K$^{-1}$ and 2.9 Wm$^{-1}$K$^{-1}$ along c-axis and a-axis respectively for the bulk MgPb$_2$Te. This ultra-low thermal conductivity is mainly attributed with very low phonon group velocities of TA and LA phonon modes due to the constituent heave elements. Likewise, we also report the length dependence thermal conductivity for the nanostructured MgPb$_2$Te. At nanometer length scales of L= 50 nm, room temperature thermal conductivity of 0.957 Wm$^{-1}$K$^{-1}$ and 1.459 Wm$^{-1}$K$^{-1}$ for the MgPb$_2$Te shows a promising nature of thermoelectric applications. This work provides an avenue to further explore this MgPb$_2$Te with point-defects and alloying etc.,

**Conflicts of Interest**

There are no conflicts of interest to declare.

**Acknowledgements**

RM and JG acknowledge support from National Science Foundation CAREER award under Award No. #1847129. We also acknowledge OU Supercomputing Center for Education and Research (OSCER) for providing computing resources for this work.

**References:**
1. Gayner, C.; Kar, K. K., Recent advances in thermoelectric materials. *Progress in Materials Science* **2016,** *83*, 330-382.
2. DiSalvo, F. J., Thermoelectric Cooling and Power Generation. *Science* **1999,** *285* (5428), 703-706.
3. Kishore, R. A.; Nozariasbmarz, A.; Poudel, B.; Sanghadasa, M.; Priya, S., Ultra-high performance wearable thermoelectric coolers with less materials. *Nature Communications* **2019,** *10* (1), 1765.
4. Gillott, M.; Jiang, L.; Riffat, S., An investigation of thermoelectric cooling devices for small-scale space conditioning applications in buildings. *International Journal of Energy Research* **2010,** *34* (9), 776-786.
5. Snyder, G. J.; Snyder, A. H., Figure of merit ZT of a thermoelectric device defined from materials properties. *Energy & Environmental Science* **2017,** *10* (11), 2280-2283.
6. Xiao, Y.; Zhao, L.-D., Seeking new, highly effective thermoelectrics. *Science* **2020,** *367* (6483), 1196-1197.
7. Bhaskar, A.; Pai, Y.-H.; Wu, W.-M.; Chang, C.-L.; Liu, C.-J., Low thermal conductivity and enhanced thermoelectric performance of nanostructured Al-doped ZnTe. *Ceramics International* **2016,** *42* (1, Part B), 1070-1076.
8. Tarannum, F.; Muthaiah, R.; Annam, R. S.; Gu, T.; Garg, J., Effect of Alignment on Enhancement of Thermal Conductivity of Polyethylene–Graphene Nanocomposites and Comparison with Effective Medium Theory. *Nanomaterials* **2020,** *10* (7), 1291.
9. Muthaiah, R.; Garg, J., Strain tuned high thermal conductivity in boron phosphide at nanometer length scales – a first-principles study. *Physical Chemistry Chemical Physics* **2020,** *22* (36), 20914-20921.
10. Muthaiah, R.; Garg, J., Temperature effects in the thermal conductivity of aligned amorphous polyethylene—A molecular dynamics study. *Journal of Applied Physics* **2018,** *124* (10), 105102.

11. Li, X.; Liu, J.; Yang, R. In *Tuning Thermal Conductivity With Mechanical Strain*, 2010 14th International Heat Transfer Conference, 2010; pp 551-558.
12. Nakamura, Y., Nanostructure design for drastic reduction of thermal conductivity while preserving high electrical conductivity. *Sci Technol Adv Mater* **2018,** *19* (1), 31-43.
13. Perumal, S.; Roychowdhury, S.; Biswas, K., Reduction of thermal conductivity through nanostructuring enhances the thermoelectric figure of merit in Ge1–xBixTe. *Inorganic Chemistry Frontiers* **2016,** *3* (1), 125-132.
14. Rounds, R.; Sarkar, B.; Alden, D.; Guo, Q.; Klump, A.; Hartmann, C.; Nagashima, T.; Kirste, R.; Franke, A.; Bickermann, M.; Kumagai, Y.; Sitar, Z.; Collazo, R., The influence of point defects on the thermal conductivity of AlN crystals. *Journal of Applied Physics* **2018,** *123* (18), 185107.
15. Dughaish, Z. H., Lead telluride as a thermoelectric material for thermoelectric power generation. *Physica B: Condensed Matter* **2002,** *322* (1), 205-223.
16. Agashe, V. V.; Sharma, B. L.; Sachar, B. K., Preparation and Properties of Lead-Telluride for Thermoelectric Generator. *IETE Journal of Research* **1964,** *10* (7), 237-240.
17. Witting, I. T.; Chasapis, T. C.; Ricci, F.; Peters, M.; Heinz, N. A.; Hautier, G.; Snyder, G. J., The Thermoelectric Properties of Bismuth Telluride. *Advanced Electronic Materials* **2019,** *5* (6), 1800904.
18. Mamur, H.; Bhuiyan, M. R. A.; Korkmaz, F.; Nil, M., A review on bismuth telluride (Bi2Te3) nanostructure for thermoelectric applications. *Renewable and Sustainable Energy Reviews* **2018,** *82*, 4159-4169.
19. Zhang, F.; Chen, C.; Yao, H.; Bai, F.; Yin, L.; Li, X.; Li, S.; Xue, W.; Wang, Y.; Cao, F.; Liu, X.; Sui, J.; Zhang, Q., High-Performance N-type Mg3Sb2 towards Thermoelectric Application near Room Temperature. *Advanced Functional Materials* **2020,** *30* (5), 1906143.
20. Li, J.; Zheng, S.; Fang, T.; Yue, L.; Zhang, S.; Lu, G., Computational prediction of a high ZT of n-type Mg3Sb2-based compounds with isotropic thermoelectric conduction performance. *Physical Chemistry Chemical Physics* **2018,** *20* (11), 7686-7693.
21. Mao, J.; Zhu, H.; Ding, Z.; Liu, Z.; Gamage, G. A.; Chen, G.; Ren, Z., High thermoelectric cooling performance of n-type $Mg_3Bi_2$-based materials. *Science* **2019**, eaax7792.
22. Peng, W.; Petretto, G.; Rignanese, G.-M.; Hautier, G.; Zevalkink, A., An Unlikely Route to Low Lattice Thermal Conductivity: Small Atoms in a Simple Layered Structure. *Joule* **2018,** *2* (9), 1879-1893.
23. Zhou, T.; Feng, Z.; Mao, J.; Jiang, J.; Zhu, H.; Singh, D. J.; Wang, C.; Ren, Z., Thermoelectric Properties of Zintl Phase YbMg2Sb2. *Chemistry of Materials* **2020,** *32* (2), 776-784.
24. Shiga, T.; Shiomi, J.; Ma, J.; Delaire, O.; RadzzDski, T.; Lusakowski, A.; Esfarjani, K.; Chen, G., Microscopic mechanism of low thermal conductivity in lead telluride. *Physical Review B* **2012,** *85*, 155203.
25. Zhang, J.; Song, L.; Iversen, B. B., Insights into the design of thermoelectric Mg3Sb2 and its analogs by combining theory and experiment. *npj Computational Materials* **2019,** *5* (1), 76.
26. Casimir, H. B. G., Note on the conduction of heat in crystals. *Physica* **1938,** *5* (6), 495-500.

27. Giannozzi, P.; Baroni, S.; Bonini, N.; Calandra, M.; Car, R.; Cavazzoni, C.; Ceresoli, D.; Chiarotti, G. L.; Cococcioni, M.; Dabo, I.; Dal Corso, A.; de Gironcoli, S.; Fabris, S.; Fratesi, G.; Gebauer, R.; Gerstmann, U.; Gougoussis, C.; Kokalj, A.; Lazzeri, M.; Martin-Samos, L.; Marzari, N.; Mauri, F.; Mazzarello, R.; Paolini, S.; Pasquarello, A.; Paulatto, L.; Sbraccia, C.; Scandolo, S.; Sclauzero, G.; Seitsonen, A. P.; Smogunov, A.; Umari, P.; Wentzcovitch, R. M., QUANTUM ESPRESSO: a modular and open-source software project for quantum simulations of materials. *Journal of Physics: Condensed Matter* **2009,** *21* (39), 395502.
28. Monkhorst, H. J.; Pack, J. D., Special points for Brillouin-zone integrations. *Physical Review B* **1976,** *13* (12), 5188-5192.
29. Chung, D. H.; Buessem, W. R., The Voigt-Reuss-Hill (VRH) Approximation and the Elastic Moduli of Polycrystalline ZnO, TiO2 (Rutile), and α-Al2O3. *Journal of Applied Physics* **1968,** *39* (6), 2777-2782.
30. Garg, J.; Bonini, N.; Kozinsky, B.; Marzari, N., Role of Disorder and Anharmonicity in the Thermal Conductivity of Silicon-Germanium Alloys: A First-Principles Study. *Physical Review Letters* **2011,** *106* (4), 045901.
31. Debernardi, A.; Baroni, S.; Molinari, E., Anharmonic Phonon Lifetimes in Semiconductors from Density-Functional Perturbation Theory. *Physical Review Letters* **1995,** *75* (9), 1819-1822.
32. Deinzer, G.; Birner, G.; Strauch, D., Ab initio calculation of the linewidth of various phonon modes in germanium and silicon. *Physical Review B* **2003,** *67* (14), 144304.
33. Paulatto, L.; Errea, I.; Calandra, M.; Mauri, F., First-principles calculations of phonon frequencies, lifetimes, and spectral functions from weak to strong anharmonicity: The example of palladium hydrides. *Physical Review B* **2015,** *91* (5), 054304.
34. Paulatto, L.; Mauri, F.; Lazzeri, M., Anharmonic properties from a generalized third-order ab initio approach: Theory and applications to graphite and graphene. *Physical Review B* **2013,** *87* (21), 214303.
35. Fugallo, G.; Lazzeri, M.; Paulatto, L.; Mauri, F., Ab initio variational approach for evaluating lattice thermal conductivity. *Physical Review B* **2013,** *88* (4), 045430.
36. Mouhat, F.; Coudert, F.-X., Necessary and sufficient elastic stability conditions in various crystal systems. *Physical Review B* **2014,** *90* (22), 224104.
37. Etxebarria, I.; Perez-Mato, J. M.; Madariaga, G., Lattice dynamics, structural stability, and phase transitions in incommensurate and commensurate ${\mathit{A}}_{2}$${\mathit{BX}}_{4}$ materials. *Physical Review B* **1992,** *46* (5), 2764-2774.
38. Ward, A.; Broido, D. A., Intrinsic phonon relaxation times from first-principles studies of the thermal conductivities of Si and Ge. *Physical Review B* **2010,** *81* (8), 085205.